\documentclass[letterpaper,12pt]{article}
\usepackage{amsmath}
\usepackage{latexsym}
\usepackage{amssymb}
\newtheorem{lem}{Lemma}
\newtheorem{thm}{Theorem}

\def\keywords{\vspace{-.3em}
    \if@twocolumn
      \small\it Keywords\/\bf---$\!$%
    \else
      \begin{center}\small\bf Keywords\end{center}\quotation\small
    \fi}
\def\endkeywords{\vspace{0.6em}\par\if@twocolumn\else\endquotation\fi
    \normalsize\rm}
\def\appendix{\par
    \setcounter{section}{0}\setcounter{subsection}{0}
    \def\thesection{\Alph{section}} \section*{Appendix}
}
\def\Tr{\mathop{\rm Tr}\nolimits}

\def\SL{\mathop{\rm SL}\nolimits}

\def\Id{\mathop{\rm I}\nolimits}

\def\rank{\mathop{\rm rank}\nolimits}

\def\complex{\mathbb{C}}

\def\Label#1{\label{#1}\ [\ #1\ ]\ }
\def\Label{\label}
\begin{document}
\bibliographystyle{IEEE}
\title{Optimal sequence of POVMs in the sense of Stein's lemma
in quantum hypothesis testing}
\author{
Masahito Hayashi\thanks{
M. Hayashi is with 
Laboratory for Mathematical Neuroscience, 
Brain Science Institute, 
RIKEN,
2-1 Hirosawa, Wako, Saitama, 351-0198, Japan 
(E-mail: masahito@brain.riken.go.jp).}}
\date{9 August 2001}

\maketitle

\begin{abstract}
In this paper, we give another proof of quantum Stein's lemma by 
calculating the information spectrum,
and study an asymptotic optimal measurement in the sense of Stein's lemma.
We propose a projection measurement characterized 
by the irreducible representation theory of the 
special linear group $\SL( {\cal H})$.
Specially, in spin 1/2 system,
it is realized by a simultaneous measurement
of the total momentum and a momentum of a specified direction.
\end{abstract}


\section{Introduction}
We propose an asymptotically optimal measurement
for simple quantum hypothesis testing.
As is mentioned the below, the quantum 
Stein's lemma is proved from Hiai-Petz result\cite{HP} and Ogawa-Nagaoka's
result\cite{ON}.
We give another proof of the quantum 
Stein's lemma from an information spectrum viewpoint.
We consider a relation between the quantum 
Stein's lemma and the measurement proposed by Hayashi\cite{Haya}.

Let ${\cal H}$ be the Hilbert space of
interest, and ${\cal S}({\cal H})$
be the set of densities on ${\cal H}$.
When we perform a measurement corresponding to 
POVM (Positive Operator Valued Measure) 
$M= \{ M_i \}$ to the system in the state $\rho$,
the data obeys the probability 
${\rm P}_{\rho}^M=
\{ {\rm P}_{\rho}^M(i) = \Tr M_i \rho\}$.
In particular, the POVM $M= \{M_i\}$ is called a PVM (Projection
Valued Measure) if any $M_i$ is a projection.
In the hypothesis testing, the testing is described by a 2-valued POVM 
$\{ M_a , M_r \}$, where $M_a$ corresponds to accept
and $M_r$ corresponds to reject.
In the sequel, an operator $A$ satisfying $0 \le A \le \Id $
is called a {\it test}
identifying it with the POVM $\{ M_a , M_r \}=
\{ A, \Id - A\}$.

We introduce the quantum $n$-i.i.d.\ condition
in order to treat an asymptotic setting.
Suppose that $n$ independent physical systems are
given in the same state $\rho$,
then the quantum state of the composite system is described 
by $\rho^{\otimes n}$ defined by
\begin{eqnarray*}
\rho^{\otimes n} :=  \underbrace{\rho \otimes \cdots \otimes \rho }_{n}
\hbox{ on } {\cal H}^{\otimes n} ,
\end{eqnarray*}
where the tensored space ${\cal H}^{\otimes n}$ is defined by
\begin{eqnarray*}
{\cal H}^{\otimes n} := 
\underbrace{{\cal H} \otimes \cdots \otimes {\cal H}}_{n} .
\end{eqnarray*}
We call this condition the quantum $n$-i.i.d.\ condition,
which is a quantum analogue of the independent-identical
distribution condition.
Under the quantum $n$-i.i.d. condition,
the equation
\begin{eqnarray*}
D(\rho^{\otimes n} \| \sigma^{\otimes n} )=
n D(\rho \| \sigma)
\end{eqnarray*}
holds, where $D(\rho \|\sigma)$ is the quantum relative entropy
$\Tr \rho(\log \rho - \log \sigma)$.

Under the quantum $n$-i.i.d. condition,
we study the hypothesis testing problem
for the null hypothesis $H_0:
\rho^{\otimes n}\in {\cal S}({\cal H}^{\otimes n})$
versus the alternative hypothesis $H_1:
\sigma^{\otimes n}\in {\cal S}({\cal H}^{\otimes n})$,
where $\rho^{\otimes n}$ and $\sigma^{\otimes n}$
are the $n$th-tensor powers of arbitrarily
given density operators
$\rho$ and $\sigma$ in ${\cal S}({\cal H})$.
In the sequel, an operator $A^n$ on ${\cal H}^{\otimes n}$
satisfying $0 \le A^n \le \Id$
or a sequence $\{ A^n \}$ of such operators, is called a {\it test}.
For a test $A^n$ the error probabilities of the first and the second are,
respectively, defined by
\begin{eqnarray*}
\alpha_n(A^n)= \Tr \rho^{\otimes n}( \Id - A^n) 
\hbox{ and }
\beta_n(A^n)= \Tr \sigma^{\otimes n} A^n. 
\end{eqnarray*}
We can understand that $\alpha_n(A^n)$
is the probability of erroneously
rejecting $\rho^{\otimes n}$
when $\rho^{\otimes n}$
is true and $\beta_n(A^n)$ is the error probability
of erroneously accepting $\rho^{\otimes n}$.
when $\rho^{\otimes n}$ is not true.
We discuss the trade-off of the two type error probabilities,
under the quantum $n$-i.i.d. condition.

The following is well-known as quantum Stein's lemma.
\begin{thm}\Label{Stein}
For any $1 \,> \epsilon \,> 0$, the equation 
\begin{eqnarray}
\lim_{n \to \infty}
\frac{1}{n} \log \beta_n^*(\epsilon)
= - D(  \rho\| \sigma)
\end{eqnarray}
holds, where 
\begin{eqnarray}
\beta_n^*(\epsilon)
:= \min \{ \beta_n(A^n)|
0 \le A^n \le \Id , \alpha_n(A^n) \le \epsilon \}.
\end{eqnarray}
\end{thm}
The part of $\ge$ was proved by Hiai-Petz \cite{HP}.
The infinite-dimensional case was proved by Petz\cite{Petz}.
The part of $\le$ is proved by Ogawa-Nagaoka\cite{ON}.
The purpose of this paper follows:
One is a construction of the testing whose 1st error probability goes
to 0 and whose 2nd error probability goes to 0 with the decreasing rate
$D(  \rho\| \sigma)$.
The other is giving another proof of Theorem \ref{Stein}
from an information spectrum method,
which is initiated by Han-Verd\'{u} \cite{HV1} and Han \cite{Han}.
An application of the information spectrum method
to quantum hypothesis testing was initiated by Nagaoka\cite{Naga,Naga2}.
An information spectrum approach to exponents in quantum hypothesis
testing 
was discussed by Nagaoka-Hayashi\cite{NH},
which can be regarded as a quantum analogue of Han \cite{Han2}.
This work was motivated by Nagaoka\cite{Naga,Naga2}.

\section{Information spectrum viewpoint for quantum hypothesis
testing}
\subsection{Information spectrum methods in classical hypothesis
testing}
We summerize the information spectrum methods in classical hypothesis
testing.
Given two general sequence of probabilities
$\vec{p}= \{p_n\}$ and
$\vec{q}=\{ q_n\}$
on the same probability sets $\{\Omega_n\}$,
we may define the general hypothesis testing problem 
with $\vec{p}= \{p_n\}$ as the null hypothesis
and $\vec{q}=\{ q_n\}$ as the alternative hypothesis.
In this situation, Any {\it classical test} 
is described by a function $A^n: \Omega_n \to [0,1]$.
This notation contains a random test.
For any test $A^n$, the error probabilities of
the first and the second are, respectively, defined by
\begin{align*}
\alpha_n(A^n):=
\sum_{\omega_n \in \Omega_n}
(1- A^n(\omega_n))p_n(\omega_n) ,
\quad
\beta_n(A^n):=
\sum_{\omega_n \in \Omega_n}
A^n(\omega_n)q_n(\omega_n) .
\end{align*}
We focus the following two quantities
\begin{align*}
B ( \vec{p} \| \vec{q})
:=&
\sup \left\{
\lambda \left|
\exists \vec{A},\quad
\lim_{n \to \infty} \alpha_n(A^n) = 0,\quad
 \limsup_{n \to \infty}
\frac{1}{n} \log \beta_n(A^n) \le -\lambda 
\right. \right\} , \\
C ( \vec{p} \| \vec{q})
:=&
\sup \left\{
\lambda \left|
\exists \vec{A},\quad
\liminf_{n \to \infty} \alpha_n(A^n) \,< 1,
\quad
\limsup_{n \to \infty}
\frac{1}{n} \log \beta_n(A^n) \le -\lambda 
\right.\right\} ,
\end{align*}
and define
\begin{align*}
\underline{D}(\vec{p}\|\vec{q})&:=
\sup \left\{ \lambda \left|
\lim_{n \to \infty}
p_n \left\{ \omega_n\left| \frac{1}{n} \log \frac{p_n(\omega_n)}{q_n(\omega_n)}
\,< \lambda \right\}\right.
= 0 \right\}\right. , \\
\overline{D}(\vec{p}\|\vec{q})&:=
\inf \left\{ \lambda \left|
\lim_{n \to \infty}
p_n \left\{ \omega_n\left| \frac{1}{n} \log \frac{p_n(\omega_n)}{q_n(\omega_n)}
\,> \lambda \right\}\right.
= 0 \right\}\right. .
\end{align*}
We have the following lemma
\begin{lem}\Label{L0}
Han\cite{Han},Verd\'{u}\cite{Verdu},Nagaoka\cite{Naga,Naga2}
We can show the relations
\begin{align}
B(\vec{p}\|\vec{q})&= \underline{D}(\vec{p}\|\vec{q})\Label{J1} \\
 C(\vec{p}\|\vec{q})&=\overline{D}(\vec{p}\|\vec{q}) \Label{J2} \\
\underline{D}(\vec{p}\|\vec{q})
&\le \overline{D}(\vec{p}\|\vec{q}).\Label{J3} 
\end{align}
\end{lem}
The equation(\ref{J1}) was proved in Chapter 4 in Han\cite{Han}.
He referred to Verd\'{u}\cite{Verdu}.
The equation(\ref{J2}) was 
pointed by Nagaoka\cite{Naga,Naga2}.
For reader's convenience, we give a proof in Appendix A.
\subsection{Information spectrum characterization of 
quantum hypothesis testing}
According to Nagaoka\cite{Naga,Naga2},
we discuss the following two quantities
\begin{align*}
B ( \vec{\rho} \| \vec{\sigma})
:=&
\sup \left\{
\lambda \left|
\exists \vec{A},
\lim_{n \to \infty} \alpha_n(A^n) = 0,
 \limsup_{n \to \infty}
\frac{1}{n} \log \beta_n(A^n) \le -\lambda 
\right. \right\} , \\
C ( \vec{\rho} \| \vec{\sigma})
:=&
\sup \left\{
\lambda \left|
\exists \vec{A},
\limsup_{n \to \infty} \alpha_n(A^n) \,< 1,
\limsup_{n \to \infty}
\frac{1}{n} \log \beta_n(A^n) \le -\lambda 
\right.\right\}  .
\end{align*}
For any sequence $\vec{M}:=\{ M^n \}$ of POVMs, we define
\begin{align*}
\underline{D}^{\vec{M}}\left(
\vec{\rho} \| \vec{\sigma} \right) 
:= 
\underline{D}\left( \left. \left\{ 
{\rm P}^{M^n}_{\rho^{\otimes n}}\right\} \right\|
\left\{ {\rm P}^{M^n}_{\sigma^{\otimes n}}\right\}\right),
\quad
\overline{D}^{\vec{M}}\left(
\vec{\rho} \| \vec{\sigma} \right) 
:= \overline{D}\left( \left. \left\{ 
{\rm P}^{M^n}_{\rho^{\otimes n}}\right\} \right\|
\left\{ {\rm P}^{M^n}_{\sigma^{\otimes n}}\right\}\right).
\end{align*}
From Lemma \ref{L0}, we have
\begin{align}
B( \vec{\rho} \| \vec{\sigma} )=\sup_{\vec{M}:\hbox{\small POVMs}} \underline{D}^{\vec{M}}\left(
\vec{\rho} \| \vec{\sigma} \right) \le
C( \vec{\rho} \| \vec{\sigma} )=\sup_{\vec{M}:\hbox{\small POVMs}} \overline{D}^{\vec{M}}\left(
\vec{\rho} \| \vec{\sigma} \right) . \Label{8.1}
\end{align}
As is proved in the latter,
the equations
\begin{align}
\sup_{\vec{M}:\hbox{\small POVMs}} \underline{D}^{\vec{M}}\left(
\vec{\rho} \| \vec{\sigma} \right) 
=
\sup_{\vec{M}:\hbox{\small PVMs}} \underline{D}^{\vec{M}}\left(
\vec{\rho} \| \vec{\sigma} \right), \quad
\sup_{\vec{M}:\hbox{\small POVMs}} \overline{D}^{\vec{M}}\left(
\vec{\rho} \| \vec{\sigma} \right) 
= \sup_{\vec{M}:\hbox{\small PVMs}} \overline{D}^{\vec{M}}\left(
\vec{\rho} \| \vec{\sigma} \right) \Label{dd1}
\end{align}
hold. In this paper, we show the equations
\begin{align}
\sup_{\vec{M}} \underline{D}^{\vec{M}}\left(
\vec{\rho} \| \vec{\sigma} \right)=
\sup_{\vec{M}} \overline{D}^{\vec{M}}\left(
\vec{\rho} \| \vec{\sigma} \right)= D (\rho \| \sigma ) , \Label{8.3}
\end{align}
which imply $B( \vec{\rho} \| \vec{\sigma} )=
C( \vec{\rho} \| \vec{\sigma} )=D (\rho \| \sigma )$, i.e.
Theorem \ref{Stein},
and construct a test $\{ A^n\}_{n=1}^{\infty}$
satisfying
\begin{align}
\lim_{n \to \infty} \alpha_n(A^n) = 0 , \quad
-\lim_{n \to \infty}\frac{1}{n} \log \beta_n(A^n)
= D ( \rho \| \sigma ) - \epsilon , \Label{2}
\end{align}
for any $\epsilon \,> 0$.
In the sequel, a test $\{A^n \}$
satisfying (\ref{2}) is called an {\it optimal test 
in the sense of Stein's lemma}.
According to Han\cite{Han}, for any $1 \,> \epsilon \,> 0$,
we can prove that the test:
\begin{itemize}
\item 
If $\frac{1}{n} \log \frac
{{\rm P}^{M^n}_{\rho^{\otimes n}}(i)}{{\rm P}^{M^n}_{\sigma^{\otimes n}}(i)} \ge 
\underline{D}^{\vec{M}}\left(
\vec{\rho} \| \vec{\sigma} \right) - \epsilon $, then
$\rho$ is accept.
\item 
If $\frac{1}{n} \log \frac
{{\rm P}^{M^n}_{\rho^{\otimes n}}(i)}{{\rm P}^{M^n}_{\sigma^{\otimes n}}(i)} \,<
\underline{D}^{\vec{M}}\left(
\vec{\rho} \| \vec{\sigma} \right) - \epsilon $, then
$\rho$ is reject.
\end{itemize}
satisfies 
\begin{align}
\lim_{n \to \infty} \alpha_n(A^n) = 0 , ~
-\lim_{n \to \infty}\frac{1}{n} \log \beta_n(A^n) 
= \underline{D}^{\vec{M}}\left(
\vec{\rho} \| \vec{\sigma} \right)
- \epsilon . \Label{65}
\end{align}
Therefore, if we can construct a sequence 
$\vec{M}:= \{ M^n \}_{n=1}^{\infty}$
of POVM satisfying 
\begin{eqnarray}
\underline{D}^{\vec{M}}\left(
\vec{\rho} \| \vec{\sigma} \right)
= D ( \rho \| \sigma ) , \Label{3}
\end{eqnarray}
then we can construct a test satisfying (\ref{2}).

In general, we have 
\begin{eqnarray}
\liminf_{n \to \infty}
\frac{1}{n}
D^{M^n}( \rho^{\otimes n} \| \sigma^{\otimes n})
\ge 
\underline{D}^{\vec{M}}\left(
\vec{\rho} \| \vec{\sigma} \right) , \quad
D ( \rho \| \sigma ) \ge D^{M} \left({\rho} \| {\sigma}\right),
\label{4}
\end{eqnarray}
where $D^M(\rho \| \sigma ):= D( {\rm P}_\rho^M\|{\rm P}_\sigma^M)$.
The second inequality (\ref{4}) can be regarded as a special
case of the monotonicity of quantum relative entropy.
Therefore, the part of $\le$ in (\ref{3}) is trivial.
We need to construct $\vec{M}$
satisfying the part of $\ge$ in (\ref{3}).
In the sequel, we call a sequence $\vec{M}$ of POVMs
an {\it optimal sequence of POVMs in the sense
of Stein's lemma}.
In the following, we prove (\ref{8.3})
from group representation viewpoint,
and construct an optimal sequence of POVMs in the sense
of Stein's lemma, which is independent of the null hypothesis
$\rho$.
In this paper, we assume that the dimension of ${\cal H}$
is finite ($k$) and the inverse $\sigma^{-1}$
of $\sigma$ exists.
\section{PVMs and fundamental inequalities}
We make some definitions for this purpose.
For any PVM $E= \{E_i\}$, we denote $\sup_i \rank E_i$ by $w(E)$.
A state $\rho$ is called {\it commutative} with 
a PVM $E(=\{ E_i \})$ on ${\cal H}$ 
if $\rho E_i = E_i \rho $ for any index $i$.
For PVMs $E (=\{ E_i \}_{i \in I}) ,F(=\{ F_j \}_{j \in J})$, 
the notation $E \le  F$ means that
for any index $i \in I$ there exists
a subset $(F/E)_i$ of the index set $J$
such that $E_i = \sum_{j \in (F/E)_i} F_j$.
For any operator  $X$, 
we denote $E(X)$ by the spectral measure of $X$
which can be regarded as a PVM.
In particular, we have $E(\sigma)= E(\log \sigma)$.
The map ${\cal E}_E$ 
with respect to a PVM $E$ is defined as:
\begin{eqnarray*}
 {\cal E}_E : \rho \mapsto \sum_{i} E_i \rho E_i , %\Label{11.4.1}
\end{eqnarray*}
which is 
an affine map from the set of states to itself.
Note that the state ${\cal E}_E(\rho)$ is commutative with a PVM $E$.
If a PVM $F=\{ F_j \}$ is commutative with a PVM $E=\{ E_i \}$, then
we can define the PVM $F \times E= \{ F_j E_i \}$,
which satisfies that $F \times E \ge E$ and $F \times E \ge F$,
and can be regarded as the simultaneous measurement of $E$ and $F$.
If a test $A$ and a PVM $M$ satisfy that $M \ge E(A)$, there
exists a classical test in the hypothesis: ${\rm P}_\rho^M$ v.s.
${\rm P}_\sigma^M$ corresponding to the test $A$.
This fact and Lemma \ref{L0} imply (\ref{dd1}).
\begin{lem}\Label{LL1}
If $\rho$ and $\sigma$ are commutative with
a PVM $E$, then the equation
\begin{align*}
\inf\left\{
\beta ( A ) \left |
\alpha (A)  \le \epsilon \right. \right\}
=\inf\left\{
\beta ( A ) \left |
\exists M : {\rm PVM }
,\quad M\ge E , 
M \ge E(A)  , 
\alpha (A)  \le \epsilon , 
w(M)=1
\right. \right\} 
\end{align*}
holds.
\end{lem}
\begin{proof}
For any $A$,
the relations $\beta({\cal E}_E(A))=\beta(A),
\alpha({\cal E}_E(A))=\alpha(A)$
hold.
Since the PVM $E({\cal E}_E(A))$ commutes with
the PVM $E$,
there exits a PVM $M$ such that 
$M \ge E, M \ge E({\cal E}_E(A))$ and $w(M)=1$.
\end{proof}
From lemma 2, we may discuss only PVMs $M$
satisfying $M \ge E$ in the above situation.
\begin{lem}\Label{lem2}
If PVMs $E,M$ satisfy that $M \ge E$
and a state $\rho$ is commutative with $E$
nd $w(E) \ge 3$, then
the inequality
\begin{align}
\Tr \rho
(\log \rho - \log {\cal E}_{M}(\rho))^2
\le 4(\log w(E))^2 \Label{62} . 
\end{align}
holds.
\end{lem}
\begin{proof}
Define $a_i :=\Tr E_i \rho E_i ,\rho_i: = \frac{1}{a_i} E_i \rho E_i$,
then the equations
$\rho = \sum_{i}a_i \rho_i,
{\cal E}_M(\rho) =  \sum_{i}a_i{\cal E}_M(\rho_i)$
hold.
Using the operator inequality
$(A+B)^2 \le 2 (A^2 B^2)$,
we have 
\begin{align*}
&\Tr \rho
(\log \rho - \log {\cal E}_{M}(\rho))^2 
=
\sum_{i} a_i 
\Tr \rho_i
(\log \rho_i - \log {\cal E}_{M}(\rho_i))^2  \\
\le &
\sup_{i} 
\Tr \rho_i
(\log \rho_i - \log {\cal E}_{M}(\rho_i))^2 
\le 
\sup_i
\Tr \rho_i
2 \left( (\log \rho_i)^2 +(\log {\cal E}_{M}(\rho_i))^2 \right) \\
= &
2\sup_i
\Tr \rho_i
(\log \rho_i)^2 
+\Tr{\cal E}_{M}(\rho_i) (\log {\cal E}_{M}(\rho_i))^2 
\le  4 \sup_i(\log \dim E_i)^2,
\end{align*}
where the last inequality follows from Lemma \ref{2jou}.
We obtain (\ref{62}).
\end{proof}
\begin{lem}\Label{2jou}
Nagaoka\cite{Naga3}, Osawa\cite{Osawa} 
The equation 
\begin{align}
& \max \left\{\left.  \sum_{i=1}^k p_i (\log p_i)^2 \right|
 p_i \ge 0, \sum_{i=1}^k p_i= 1\right\} \nonumber \\
= &
\left\{
\begin{array}{cc}
 (\log k)^2 & \hbox{ if } k \ge 3 \\
\frac{1-\sqrt{1-\frac{4}{e^2}}}{2}
\left( \log \frac{1-\sqrt{1-\frac{4}{e^2}}}{2}\right)^2
+
\frac{1+\sqrt{1-\frac{4}{e^2}}}{2}
\left( \log \frac{1+\sqrt{1-\frac{4}{e^2}}}{2}\right)^2
& \hbox{ if } k = 2
\end{array}
\right.
.\Label{22}
\end{align}
holds.
\end{lem}
Its proof is given in Appendix B.
\begin{lem}\Label{c1}
Let $k$ be the dimension of ${\cal H}$.
For any state $\rho \in {\cal S}({\cal H})$
and any PVM $M$,
the inequality 
$\rho \le {\cal E}_M( \rho ) k$ holds.
\end{lem}
\begin{proof}
It is sufficient to prove 
the inequality for any pure state $| \phi \rangle \langle \phi |$.
We have 
\begin{align*}
\left\langle \psi \left| \left(
{\cal E}_M( | \phi \rangle \langle \phi | ) k 
- | \phi \rangle \langle \phi |  
\right)\right| \psi \right\rangle 
=  k \sum_{i=1}^k
 \langle \psi |M_i | \phi \rangle \langle \phi |M_i  | \psi \rangle
- \left| \sum_{i=1}^k \langle \psi |M_i | \phi \rangle \right|^2 \ge 0 ,
\end{align*}
for any $\psi \in {\cal H}$,
where the inequality follows from 
Schwarz' inequality about vectors\par
\noindent
$\{ \langle \psi |M_i | \phi \rangle 
\}_{i=1}^k, \{1 \}_{i=1}^k$.
The proof is completed.
\end{proof}
\begin{lem}\Label{c2}\rm
Let $\rho$ be a state commuting the PVM $E$.
If PVM $M$ satisfies that $M \ge E$,
the inequality 
$\rho \le {\cal E}_M( \rho ) w(E)$ holds.
Since the map $u \to - u^{-t} ~(0 \,< t \le 1)$ is 
an operator monotone function in $(0,\infty)$,
when $\rho^{-1}$ is bounded,
the inequality $w(E)^{t}\rho^{-t} \ge \left({\cal E}_M( \rho )\right)^{-t} $
holds.
\end{lem}
\begin{proof}
It is immediate from Lemma \ref{c1}.
\end{proof}
\section{Relation between $\rho^{\otimes n},\sigma^{\otimes n}$ 
and group representation}\Label{s4}
In this section, we discuss the quantum i.i.d. condition
from a group theoretical viewpoint.
In \S \ref{s31}, 
we consider the relation between 
irreducible representations and PVMs.
In \S \ref{s32}, 
we discuss the quantum i.i.d. condition and PVMs 
from a theoretical viewpoint.
\subsection{group representation and its irreducible decomposition}\Label{s31}
Let $V$ be a finite dimensional vector space 
over the complex numbers $\complex$.
A map $\pi$ from a group $G$ to the generalized linear group
of a vector space $V$ is called a {\it representation} on $V$
if the map $\pi$ is homomorphism i.e. 
$\pi( g_1 ) \pi (g_2) = \pi ( g_1 g_2 ), ~ \forall g_1 , g_2 \in G$.
A subspace $W$ of $V$ is called {\it invariant} with respect to 
a representation $\pi$ if the vector $\pi (g) w$ belongs to the
subspace $W$ for any vector $w \in W$ and any element $g \in G$.
A representation $\pi$ is called {\it irreducible} if
there is no proper nonzero invariant subspace of $V$
with respect to $\pi$.
Let $\pi_1$ and $\pi_2$ be representations of a group $G$ on 
$V_1$ and $V_2$, respectively.
The {\it tensored} representation $\pi_1 \otimes \pi_2$ of $G$ on $V_1 \otimes V_2$
is defined as
$(\pi_1 \otimes \pi_2) (g) 
= \pi_1  (g) \otimes \pi_2 (g) $,
and the {\it direct sum}
 representation $\pi_1 \oplus \pi_2$ of $G$ on $V_1 \oplus
 V_2$ is also defined as
$(\pi_1 \oplus \pi_2) (g) 
= \pi_1  (g) \oplus \pi_2 (g) $.

In the following,
we treat a representation $\pi$ of a group $G$
on a finite-dimensional Hilbert space ${\cal H}$;
The following facts is crucial in the later arguments.
There exists an irreducible decomposition ${\cal H}=
{\cal H}_1 \oplus \cdots \oplus {\cal H}_l$
such that the irreducible components
are orthogonal to one another
if for any element $g \in G$ 
there exists an element $g^* \in G$ such that
$\pi(g)^*= \pi (g^*)$ where $\pi(g)^*$ denotes the adjoint of 
the linear map $\pi(g)$.
We can regard the irreducible decomposition ${\cal H}=
{\cal H}_1 \oplus \cdots \oplus {\cal H}_l$
as the PVM
$\{ P_{{\cal H}_i} \}_{i=1}^{l}$, where $P_{{\cal H}_i}$ denotes
the projection to ${\cal H}_i$.
If two representations $\pi_1,\pi_2$ satisfy the preceding condition,
then the tensored representation $\pi_1 \otimes \pi_2$, also, 
satisfies it.
Note that, in general,
an irreducible decomposition of a representation satisfying the
preceding condition is not unique.
In other words, we cannot uniquely define the PVM from such a representation.

\subsection{Relation between the tensored representation and PVMs}\Label{s32}
Let the dimension of the Hilbert space ${\cal H}$ be $k$.
Concerning the natural representation $\pi_{\SL({\cal H})}$ of 
the special linear group $\SL({\cal H})$ on ${\cal H}$,
we consider its $n$-th tensored representation
$\pi_{\SL({\cal H})}^{\otimes n}
:= \underbrace{\pi_{\SL({\cal H})} \otimes \cdots 
\otimes \pi_{\SL({\cal H})}}_n$ on the tensored space ${\cal H}^{\otimes n}$.
For any element $g \in \SL({\cal H})$,
the relation $\pi_{\SL({\cal H})}(g)^*=
\pi_{\SL({\cal H})}(g^*)$ holds where the element $g^* \in \SL({\cal
  H})$ denotes the adjoint matrix of the matrix $g$.
Consequently, there exists an irreducible decomposition 
of $\pi_{\SL({\cal H})}^{\otimes n}$ regarded as a PVM
and we denote the set of such PVMs by $Ir^{\otimes n}$.
      
From the Weyl's dimension formula 
((7.1.8) or (7.1.17) in Goodman-Wallch\cite{GW}),
the $n$-th symmetric tensored space is
the maximum-dimensional space in 
the irreducible subspaces with respect to the $n$-th tensored representation 
$\pi_{\SL({\cal H})}^{\otimes n}$.
Its dimension equals the repeated combination $~_{k}H_n$
evaluated by 
$~_kH_{n} =  {n+k-1 \choose k-1} = {n+k-1 \choose n}  
=~_{n+1}H_{k-1}\le (n+1)^{k-1} $.
Thus, any element $ E^n\in Ir^{\otimes n}$ satisfies that
$w( E^n ) \le (n+1)^{k-1}$.% \Label{s4.2}.
\begin{lem}\Label{thm2}
A PVM $E^n \in Ir^{\otimes n}$ 
is commutative with the $n$-th tensored state $\rho^{\otimes n}$
of any state $\rho$ on ${\cal H}$.
\end{lem}
\begin{proof}
If $\det \rho \neq 0$,
then this lemma is trivial from the fact that $\det(\rho)^{-1}
\rho \in \SL({\cal H})$.
If $\det \rho = 0$,
there exists a sequence $\{ \rho_i \}_{i=1}^{\infty}$ 
such that $\det \rho_i \neq 0$ and 
$\rho_i \to \rho$ as $i \to \infty$.
We have 
$\rho_i^{\otimes n} \to \rho^{\otimes n}$ as $i \to \infty$.
Because a PVM $E^n \in Ir^{\otimes n}$ 
is commutative with $\rho_i^{\otimes n}$,
it is, also, commutative with $\rho^{\otimes n}$.
\end{proof}

\section{Proof of $D( \rho \|\sigma) \ge 
\overline{D}^{\vec{M}}( \vec{\rho} \| \vec{\sigma})$}
Assume that $\sigma^{-1}$ exists.
States $\sigma^{\otimes n}$ and $\rho^{\otimes n}$ are commutative
with the PVM $E^n \in Ir^{\otimes n}$. 
From Lemma \ref{LL1},
We may treat only a PVM satisfying that $M^n \ge E^n$, $w(M^n)=1$.
Our main point is the asymptotic behavior 
of the variable $\frac{1}{n}\log
\frac{{\rm P}_{\rho^{\otimes n}}^{M^n}}
{
{\rm P}_{\sigma^{\otimes n}}^{M^n}}$ under the probability distribution
${\rm P}_{\rho^{\otimes n}}^{M^n}$.
We have 
\begin{align*}
&\sum_{i}{\rm P}_{\rho^{\otimes n}}^{M^n}(i)\left(
\frac{1}{n}\log {\rm P}_{\rho^{\otimes n}}^{M^n}(i)
- \Tr \rho \log \rho \right)^2 \\
=&
\Tr {\cal E}_{M^n}(\rho^{\otimes n}) \left(
\frac{1}{n}
\log {\cal E}_{M^n}(\rho^{\otimes n}) -  
\Tr \rho \log \rho \right)^2
= \Tr \rho^{\otimes n} \left(
\frac{1}{n}\log {\cal E}_{M^n}(\rho^{\otimes n}) -  
\frac{1}{n}\Tr \rho \log \rho \right)^2\\
\le & 2 \Tr \rho ^{\otimes n}
\left(\frac{1}{n}\log {\cal E}_{M^n}(\rho^{\otimes n}) - 
\frac{1}{n} \log \rho^{\otimes n} \right)^2
+ 2 \Tr \rho^{\otimes n}\left(
\frac{1}{n}\log \rho^{\otimes n} -  \Tr \rho \log \rho \right)^2
 \\
\le& 8 \left(\frac{ (k-1)\log ( n+1) }{n}\right)^2 
+ 2 \Tr \rho^{\otimes n}
\left(\frac{1}{n}(\log \rho)^{(n)} 
- \Tr \rho \log \rho \right)^2 ,
\end{align*}
where the last inequality follows from 
Lemma \ref{lem2} and Lemma \ref{thm2}.
The second term goes to $0$.
Thus, the variable $\frac{1}{n}\log {\rm P}_{\rho^{\otimes n}}^{M^n}$ 
converges to $\Tr \rho \log \rho$ in probability.
Next, we discuss the asymptotic behavior of 
the variable $\frac{1}{n}\log {\rm P}_{\sigma^{\otimes n}}^{M^n}$.
From Markov inequality,
we have 
\begin{align*}
p \{ X \ge a \} 
\le e^{- \Lambda(X,p,a)} , \quad
\Lambda(X,p,a):= \sup_{0 \le t \le 1} 
\left(a t - \log \int e^{t X(\omega)} p(\,d \omega) \right) .
\end{align*}
We can calculate 
\begin{align*}
& \Lambda\left(- \log {\rm P}_{\sigma^{\otimes n}}^{M^n},
{\rm P}_{\rho^{\otimes n}}^{M^n},a n \right) 
= \sup_{0 \le t \le 1} a n t - \log \Tr 
\left({\cal E}_{M^n}(\rho^{\otimes n})
\left({\cal E}_{M^n}(\sigma^{\otimes n})\right)^{-t} \right) \\
=& \sup_{0 \le t \le 1} a n t - \log \Tr 
\left(\rho^{\otimes n}
\left({\cal E}_{M^n}(\sigma^{\otimes n})\right)^{-t} \right) 
\ge   \sup_{0 \le t \le 1} a n t - \left(  t \log w(E^n)  
+ \log \Tr \rho^{\otimes n} \left(\sigma^{\otimes n}\right)^{-t} \right)\\
= & \sup_{0 \le t \le 1}  n \left( a t - t \frac{\log w(E^n)  }{n}
- \log \Tr \rho \sigma^{-t} \right),
\end{align*}
where the inequality $\ge$ follows from Lemma \ref{c2}.
If $a \,> - \Tr \rho \log \sigma$, then the inequality \par\noindent
$\displaystyle
\lim_{n \to \infty}
\sup_{0 \le t \le 1} \left( a t - t \frac{(k+1)\log (n+1)  }{n}
- \log \Tr \rho \sigma^{-t} \right) \,> 0$ holds.
Thus, the inequality 
\begin{align*}
- \Tr \rho \log \sigma
\ge 
\inf\left\{
\lambda \left|
\lim_{n \to \infty} {\rm P}_{\rho^{\otimes n}}^{M^n}\left\{
-\frac{1}{n}\log {\rm P}_{\sigma^{\otimes n}}^{M^n} \,> \lambda \right\}
=0 \right. \right\} 
\end{align*}
holds. Therefore, we  obtain
\begin{align}
D( \rho \| \sigma )
\ge 
\overline{D}^{\vec{M}}( \vec{\rho} \| \vec{\sigma} ).
\Label{8.6}
\end{align}
\section{Optimal sequence of POVMs in the sense of Stein's lemma}
From the above discussion,
a sequence $\vec{M}$ of PVMs satisfies (\ref{3})
iff the variables $- \frac{1}{n}\log {\rm P}_{\sigma^{\otimes n}}^{M^n}$ 
converges to $- \Tr \rho \log \sigma$ in probability.
If $M^n$ is commutative with $\sigma^{\otimes n}$ and satisfies that
$M^n \ge E^n $, $w(M^n)=1$ for a PVM $E^n \in Ir^{\otimes n}$,
the equations
\begin{align}
\sum_{i}{\rm P}_{\rho^{\otimes n}}^{M^n}(i)\left|
\frac{1}{n}\log {\rm P}_{\sigma^{\otimes n}}^{M^n}(i)
- \Tr \rho \log \sigma \right| 
=&
\Tr {\cal E}_{M^n}(\rho^{\otimes n})
\left|
\frac{1}{n}\log {\cal E}_{M^n}(\sigma^{\otimes n})
- \Tr \rho \log \sigma \right| \nonumber \\
=
\Tr \rho^{\otimes n} \left|
\frac{1}{n}
\log \sigma^{\otimes n} -  \Tr \rho \log \sigma \right| 
=&
\Tr \rho^{\otimes n} \left|
\frac{1}{n}
(\log \sigma)^{(n)} -  \Tr \rho \log \sigma \right| \Label{dd}
\end{align}
hold.
The PVM $E^n\times E({\sigma^{\otimes n}})$ is an example
of such a PVM.
The equation (\ref{dd}) implies that the variable
$\frac{1}{n}\log {\rm P}_{\sigma^{\otimes n}}^{M^n}$ converges to 
$\Tr \rho \log \sigma$ in probability.
Therefore, it satisfies (\ref{3}).
The equation (\ref{8.3}) follows from (\ref{8.6}) and 
the existence of a sequence of PVM satisfying (\ref{3}).
This PVM coincides the PVM proposed by Hayashi\cite{Haya}.

In particular, in spin 1/2 system,
$E^n \times E({\sigma^{\otimes n}})$ 
can be regarded as a simultaneous measurement 
of the total momentum and a momentum of the specified direction.

\section{Conclusion}
We give another proof of the quantum Stein's lemma
by using group representational method in the finite-dimensional case.
In the preceding proof, the direct part and the converse part
are proved in a different way.
In this paper, using an information spectrum method,
we discuss both of them from an unified viewpoint, and
prove the direct part from an equivalent condition
for the inequality corresponding to the converse part.

\section*{Appendix A: Proof of Lemma \ref{L0}}
We simplify $\underline{D}(\vec{p}\|\vec{q})$ and 
$\overline{D}(\vec{p}\|\vec{q})$ by $\underline{D}$ and $\overline{D}$,
respectively.
The inequality (\ref{J3}) is trivial.\par
{\it Direct part of (\ref{J1}):}
Define the set $ S_n(\lambda)$ by
\begin{eqnarray}
S_n(\lambda):=
\left\{ \omega_n\left|
\frac{1}{n}\log \frac{p_n(\omega_n)}{q_n(\omega_n)}
\ge \lambda\right\}\right.  .
\end{eqnarray}
and the test $A^n(\lambda)$ by the test function 
$1_{S_n(\lambda)}$.
For any $\epsilon \,> 0$,
we have
\begin{align*}
\alpha_n(A^n(\underline{D}-\epsilon))
= p_n (S_n(\underline{D}-\epsilon)^c)
=
p_n 
\left\{ \omega_n\left| \frac{1}{n} \log \frac{p_n(\omega_n)}{q_n(\omega_n)}
\,<\underline{D}-\epsilon
 \right\}\right. \to 0
\end{align*}
and
\begin{align*}
\beta_n(A^n(\underline{D}-\epsilon))
&= q_n
\left\{ \omega_n\left| \frac{1}{n} \log \frac{p_n(\omega_n)}{q_n(\omega_n)}
\ge\underline{D}-\epsilon
 \right\}\right.\\
&\le
e^{-n (\underline{D}-\epsilon)} p_n
\left\{ \omega_n\left| \frac{1}{n} \log \frac{p_n(\omega_n)}{q_n(\omega_n)}
\ge\underline{D}-\epsilon
 \right\}\right.
\le e^{-n (\underline{D}-\epsilon)}.
\end{align*}
Thus, 
\begin{align*}
\limsup_{n \to \infty}
\frac{1}{n}
\log \beta_n(A^n(\underline{D}-\epsilon))
\le
- (\underline{D}-\epsilon).
\end{align*}

{\it Direct part of (\ref{J2}):}
Note that
\begin{align*}
\overline{D}=
\sup \left\{ \lambda \left|
\liminf_{n \to \infty}
p_n \left\{ \omega_n\left| \frac{1}{n} \log \frac{p_n(\omega_n)}{q_n(\omega_n)}
\le \lambda \right\}\right.
\,< 1 \right\}\right. .
\end{align*}
For any $\epsilon \,> 0$,
similarly, 
we have
\begin{align*}
\liminf_{n \to \infty}
\alpha_n(A^n(\overline{D}-\epsilon))
&=
\liminf_{n \to \infty} p_n 
\left\{ \omega_n\left| \frac{1}{n} \log \frac{p_n(\omega_n)}{q_n(\omega_n)}
\,<\overline{D}-\epsilon
 \right\}\right. \,< 1 \\
\beta_n(A^n(\overline{D}-\epsilon))
&\le e^{-n (\overline{D}-\epsilon)}.
\end{align*}
Thus, 
\begin{align*}
\liminf_{n \to \infty}
\frac{1}{n}
\log \beta_n(A^n(\overline{D}-\epsilon))
\le
- (\overline{D}-\epsilon).
\end{align*}

{\it Converse part of (\ref{J1}):}
Assume that $\alpha_n(A^n) \to 0 $ as $n \to \infty$
and 
\begin{align*}
\limsup_{n \to \infty}
\frac{1}{n}\log \beta_n(A^n)= - R.
\end{align*}
For any $\epsilon \, > 0$,
from Neyman-Pearson lemma,
the inequality
\begin{align}
\alpha_n(A^n(R-\epsilon))+ e^{n(R-\epsilon)}\beta_n(A^n(R-\epsilon)) 
\le \alpha_n(A^n) + e^{n(R-\epsilon)}\beta_n(A^n) \Label{NP}
\end{align}
holds.
Since the RHS goes to $0$ and
$e^{n(R-\epsilon)}\beta_n(A^n(R-\epsilon)) \ge 0$,
the relation \begin{align*}
p_n \left\{ \omega_n\left| \frac{1}{n} \log \frac{p_n(\omega_n)}{q_n(\omega_n)}
\,<R -\epsilon
 \right\}\right.
= \alpha_n(A^n(R-\epsilon))
\to 0
\end{align*}
holds. It implies that $R - \epsilon \,< \underline{D}$.

{\it Converse part of (\ref{J2}):}
Assume that $\liminf_{n \to \infty}\alpha_n(A^n) \,< 1 $
and 
\begin{align}
\limsup_{n \to \infty}
\frac{1}{n}\log \beta_n(A^n)= - R. \Label{J22}
\end{align}
For any $\epsilon \, > 0$,
from (\ref{NP}) and (\ref{J22}),
we have
\begin{align*}
\liminf_{n \to \infty}
p_n \left\{ \omega_n\left| \frac{1}{n} \log \frac{p_n(\omega_n)}{q_n(\omega_n)}
\,<R -\epsilon
 \right\}\right.
= \liminf_{n \to \infty}\alpha_n(A^n(R-\epsilon))
\le \liminf_{n \to \infty}\alpha_n(A^n)\,< 1.
\end{align*}
It implies that
$R - \epsilon \,< \overline{D}$.

\section*{Appendix B: Proof of Lemma \ref{2jou}}
In the cases $k=2,3$, the equation
(\ref{22}) is cheked by a calculation.
Now, we prove (\ref{22}) by induction
in the case $k \ge 4$.
Let $a_k$ be the RHS of (\ref{22}).
The inequality $a_k \ge (\log k)^2$ is trivial.
From the assumption of the induction,
if $a_k=   \sum_{i=1}^k p_i (\log p_i)^2$,
then $p_i \,> 0~(i=1, \ldots , k)$.
Using Lagrange multiplier method,
we have 
$(\log p_i )^2+ 2 \log p_i - \lambda' =0$,
where $\lambda'$ is the Lagrange multiplier.
The solution is written by 
$\log p_i= -1 \pm \lambda$,
where $\lambda:= \sqrt{1+ \lambda'}$.
Without loss of generality,
we ca assume that 
there exists $0\le r \le k$ such that
\begin{eqnarray*}
\log p_i = \left\{
\begin{array}{cc}
-1 + \lambda & \hbox{ if } r \ge i \\
-1 - \lambda & \hbox{ if } r \,< i 
\end{array}\right.
.
\end{eqnarray*}
Since, $\sum_i p_i =1$,
we have 
\begin{eqnarray*}
1= r e^{-1+\lambda} + (k-r) e^{-1-\lambda},
\end{eqnarray*}
which is equivalent with the 
quadratic equation
\begin{eqnarray*}
rx^2 - e x + k -r =0,
\end{eqnarray*}
where $x:=e^{\lambda}$.
Since the discriminant is greater than $0$,
we have 
\begin{eqnarray*}
e^2 - 4r(k-r) \ge 0,
\end{eqnarray*}
which is solved as:
\begin{eqnarray}
r \le \frac{k- \sqrt{k^2-e^2}}{2},
 \frac{k+ \sqrt{k^2-e^2}}{2} \le r.
\Label{81}
\end{eqnarray}
The function $c(x):=
\frac{x-\sqrt{x^2-e^2}}{2}$
is monotone decreasing in $(e,\infty)$,
and $c(4) \,< 1$.
Thus, the condition (\ref{81})
implies that $r= 0$ or $k$.
Thus, we have $p_i= 1/k$, i.e. (\ref{22}).

\section*{Acknowledgment}
The author wishes 
to thank Professor H. Nagaoka, Mr. S. Osawa 
and Dr. T. Ogawa for useful comments.

\end{document}